\documentclass[preprint,showpacs,preprintnumbers,amsmath,amssymb]{revtex4}

\usepackage{graphicx}
\usepackage{dcolumn}
\usepackage{bm}

\begin{document}

\title{Nuclear multifragmentation within the framework of different
statistical ensembles}

\author{C.E.\ Aguiar}
\author{R.\ Donangelo}
\author{S.R.\ Souza}%
\affiliation{Instituto de Fisica, Universidade Federal do Rio de Janeiro\\
Cidade Universit\'aria, CP 68528, 21941-972 Rio de Janeiro, Brazil}

\date{\today}

\begin{abstract}
The sensitivity of the Statistical Multifragmentation Model to the
underlying statistical assumptions is investigated.
We concentrate on its micro-canonical, canonical, and isobaric formulations.
As far as average values are concerned, our results reveal that all the
ensembles make very similar predictions, as long as the relevant
macroscopic variables (such as temperature, excitation energy and
breakup volume) are the same in all statistical ensembles.
It also turns out that the multiplicity dependence of the breakup
volume in the micro-canonical version of the model mimics
a system at (approximately) constant pressure, at least in the plateau region
of the caloric curve.
However, in contrast to average values, our results suggest that the
distributions of physical observables are quite sensitive to the
statistical assumptions.
This finding may help deciding which hypothesis corresponds to the
best picture for the freeze-out stage.
\end{abstract}

\maketitle

\begin{section}{Introduction}
\label{sec:intro}
The thermodynamics of the multifragment emission in nuclear reactions
has been intensively investigated  both experimentally
\cite{Ma2005,bimodal2005,Elliott2003,DAgostino2002,
theNatureOfPhaseTransition2001} and
theoretically \cite{BettyPhysRep2005,LGMoretto2005,ChomazLG2005,
FreezeOut2005,CampiHC2005,LopezEntropyCC,
MorettoLV2003,MorettoNegativeHeatCapacity2002,orderParameterWagner,
Elliott2000,Chomaz2000,DasGuptaCanonicalGCanonical,dasGuptaCp,dasGuptaCC,
GrossPhysRep1997,BondorfPhysRep}.
Particularly, the qualitative aspects of the nuclear caloric curve, such as the
existence of a plateau, has been strongly debated
\cite{Ma2005,DAgostino2002,theNatureOfPhaseTransition2001,ccCampi1996,
ccNatowitz2002,ccRuangma2002,reviewBetty2001,ccPeter1998,ccKwiatkowski1998,
cc1998,ccmsu1998,ccHauger1997,ccMa1997,ccMoretto1996,ccgsi1995}.
The controversy related to the determination of this quantity
is mainly due to the great difficulties involved in the measurement
of the excitation energy and the breakup temperature of the fragmenting
system \cite{reviewBetty2001}.
Therefore, the existence of a liquid-gas phase transition in nuclear
matter could not be clearly established or ruled out yet.
Furthermore, several statistical calculations, supported by experimental
analyses \cite{DAgostino2002}, suggest that the heat
capacity at constant pressure, $C_p$, can assume negative values
under certain conditions \cite{smm2,Chomaz2000,GrossPhysRep1997,dasGuptaCp,
Elliott2000,IsobaricShlomo2004} whereas other considerations have led to
opposite conclusions \cite{CampiHC2005,MorettoNegativeHeatCapacity2002}.

Theoretical aspects related to the smallness of nuclear systems have also been
discussed and could influence conclusions drawn from statistical
calculations based upon different constraints \cite{ChomazLG2005}.
More specifically, the equivalence of statistical ensembles is
expected to be valid only for very large systems, {\it i.e.} in the
thermodynamical limit.
Of course, this condition is not fulfilled in the reactions we
discuss here.
In spite of this, statistical approaches are 
very successul in quantitatively describing many features of 
nuclear multifragmentation \cite{BondorfPhysRep,GrossPhysRep1997}.

In this work, we compare the predictions of the Statistical
Multifragmentation Model (SMM) \cite{smm1,smm2,smm3},
concentrating on its micro-canonical, canonical and isobaric versions.
By comparing them, one actually confronts
two different pictures for the freeze-out stage.
In the canonical and in the micro-canonical frameworks, the break-up volume is
fixed, so that its properties are dictated only the by range of the
nuclear forces.
The underlying assumption in the isobaric ensemble
is that the freeze-out stage is associated with a given collision
rate and, therefore, the break-up volume is allowed to fluctuate from
one microstate to another.
Since the freeze-out configuration can be rather different in both cases,
one expects to find characteristic fingerprints for some observables.

In sect.\ \ref{sec:model} we give a detailed review of the different
versions of the SMM used in this work and, in particular, of
the isobaric ensemble.
The comparison among their predictions is discussed in
sect.\ \ref{sec:comparison}.
We conclude in sect.\ \ref{sec:conclusion} with a brief summary and
with suggestions for experimental analysis.

\end{section}

\begin{section}{The model}
\label{sec:model}
The SMM is based on a scenario in which a hot and compressed source expands,
undergoing a prompt statistical breakup at a density smaller than that of
normal nuclear matter \cite{smm1}.
The different fragmentation modes $\{f\}$ are weighed according to their
statistical factors.
The latter depend on the statistical ensemble adopted
\cite{smm1,BotvinaGC,BondorfPhysRep}, although the essential ingredients
of the model do not vary.
We present below the three statistical versions used in this work.

\begin{subsection}{The Microcanonical Ensemble}
\label{sec:micro}
In the microcanonical version \cite{smm1,smm2,smm3}, each
fragmentation mode $f$ of an excited source with mass and atomic numbers
$A_0$ and $Z_0$, respectively, must strictly obey the following mass,
charge and energy constraints:

\begin{equation}
A_0=\sum_{\{AZ\}}N_{AZ}^f A\;,
\label{eq:masscons}
\end{equation}

\begin{equation}
Z_0=\sum_{\{AZ\}}N_{AZ}^f Z\;,
\label{eq:chargecons}
\end{equation}

\noindent
and

\begin{equation}
-B_{A_0,Z_0} + E^*=C_C\frac{Z^2_0}{A^{1/3}_0}
\left(\frac{V_0}{V}\right)^{1/3}+\sum_{\{A,Z\}}N_{A,Z}^f E_{A,Z}
\label{eq:econs}
\end{equation}

\noindent
where $N_{A,Z}^f$ represents, within a fragmentation mode $f$, the multiplicity
of a fragment with mass and atomic numbers $A$ and $Z$, respectively,
$B_{A_0,Z_0}$ stands for the binding energy of the source,
$E^*$ is the excitation energy deposited into the system and $T$ its
breakup temperature.
The first term on the right hand side of the equation above corresponds to the
Coulomb energy of a homogeneously charged sphere at breakup volume $V$,
whereas $V_0$ is the source's volume at normal nuclear density.
The fragment energy $E_{A,Z}(T,V)$:

\begin{equation}
E_{A,Z}(T,V)=-B_{A,Z}+E^K_{A,Z}(T)+E^*_{A,Z}(T)+E^C_{A,Z}(V)
\label{eq:eaz}
\end{equation}

\noindent
has contributions from its binding energy $B_{A,Z}$, translational motion
$E^K_{A,Z}(T)$, internal excitation energy $E^*_{A,Z}(T)$, and from the
remaining Coulomb terms, $E^C_{A,Z}(V)$, which contribute to the total
Coulomb energy of the system in the Wigner-Seitz
approximation \cite{WignerSeitz,smm1}.
Throughout this paper, we use the Liquid Drop formula adopted in
ref. \cite{smm3}:

\begin{eqnarray}
B_{A,Z} & = & w_0A-\beta_0A^{2/3}-C_C\frac{Z^2}{A^{1/3}}\nonumber\\
&-& K_{\rm asym}\frac{(A-2Z)^2}{A}/\left[1+\frac{9}{4}
\frac{K_{\rm asym}}{Q_{\rm asym}A^{1/3}}\right]
\label{eq:ldf}
\end{eqnarray}

\noindent
for all fragments with $A>4$.
Empirical binding energies are used for lighter nuclei.
The parameters entering in the expression above are $w_0=16.0$~MeV,
$\beta_0=18.0$~MeV, $C_C=0.737$~MeV, $K_{\rm asym}=30.0$~MeV,
and $Q_{\rm asym}=35.0$~MeV.
The remaining energy terms in Eq.\ (\ref{eq:eaz}) read:

\begin{equation}
E^K_{A,Z}(T)=\frac{3}{2}T\;,
\label{eq:ek}
\end{equation}

\begin{equation}
E^*_{A,Z}(T)=\frac{T^2}{\epsilon_0}A
+\left(\beta(T)-T\frac{d\beta}{dT}-\beta_0\right)A^{2/3}\;,
\label{eq:eex}
\end{equation}

\noindent

\begin{equation}
\beta(T)=\left[\frac{T^2_c-T^2}{T^2_c+T^2}\right]^{5/4}\;,
\label{eq:beta}
\end{equation}

and

\begin{equation}
E^C_{A,Z}(V)=-C_C\frac{Z^2}{A^{1/3}}\left(\frac{V_0}{V}\right)^{1/3}\;,
\label{eq:ec}
\end{equation}

\noindent
where $\epsilon_0=16.0$~MeV and $T_c=18.0$~MeV is the critical
temperature above which the surface tension vanishes.
Light nuclei, $A<5$, are treated as particles without internal degrees of
freedom. Therefore, they only contribute to the total energy through
the binding and kinetic energies, except for alpha particles in
which case one retains the bulk term in Eq. (\ref{eq:eex}) to account
for its particle-stable excited states.

The energy conservation stated by Eq.\ (\ref{eq:econs}) allows one to
determine the microcanonical temperature $T_f$ for each fragmentation mode
$f$.
Therefore, $T_f$ fluctuates from one partition to the other.
The statistical weight $w_f$ of a fragmentation mode is given by its
corresponding number of states:

\begin{equation}
w_f=\exp\left(S_f\right)\;,
\label{weightMC}
\end{equation}

\noindent
where $S_f$ stands for the entropy, which is obtained from the sum of the
contributions due to each fragment.
It is related to the total energy $E$ and the Helmholtz free energy $F$
through the standard thermodynamical expression:

\begin{equation}
F=E-TS
\label{eq:fe}
\end{equation}

\noindent
and

\begin{equation}
S(T,V)=-\frac{\partial F}{\partial T}(T,V)\;.
\label{eq:smc}
\end{equation}

\noindent
The Helmholtz free energy for a fragmentation mode $f$ can be written as:

\begin{eqnarray}
F_f(T,V)&=& F^*_f(T)+F_f^{\rm trans}(T,V)+F_f^C(V)\nonumber\\
&-&\sum_{\{A,Z\}}N_{A,Z}^fB_{A,Z}
\label{eq:helmfe}
\end{eqnarray}

\noindent
where

\begin{eqnarray}
F^*_f(T)&=&\sum_{\{A,Z\}}N_{A,Z}^fF^*_{A,Z}(T)\nonumber\\
&=&\sum_{\{A,Z\}}N_{A,Z}^f\left[-\frac{T^2}{\epsilon_0}A+
      \left[\beta(T)-\beta_0\right]A^{2/3}\right]\;,
\label{eq:feex}
\end{eqnarray}

\begin{eqnarray}
F_f^{\rm trans}(T,V)=T\sum_{\{A,Z\}}\Big[
&-&N_{A,Z}^f\log\left(\frac{g_{A,Z}V_{\rm free}}{\lambda_T^3}\right)\nonumber\\
&+&\log(N_{A,Z}^f!)\Big]\;,
\label{eq:fetrans}
\end{eqnarray}

\noindent

\begin{equation}
F_C^f(V)=a_C^{f}\left(\frac{V_0}{V}\right)^{1/3}\;,
\label{eq:fec}
\end{equation}

\noindent
and

\begin{equation}
a_C^{f}=C_C\left[\frac{Z_0^2}{A_0^{1/3}}
      -\sum_{\{A,Z\}}N_{A,Z}^f\frac{Z^2}{A^{1/3}}\right]\;.
\label{eq:ccnaz}
\end{equation}

\noindent
In the above equations, $g_{A,Z}$ stands for the the spin degeneracy factor
of the fragment.
It is assumed to be equal to unity for fragments which
have internal degrees of freedom whereas empirical ground state values
are used for the others.
The free volume is related to $V$ by:

\begin{equation}
V_{\rm free}=V-V_0=V_0\chi,\;\;\; 0\le\chi<\infty
\label{eq:vfree}
\end{equation}

\noindent
and $\lambda_T=\sqrt{2\pi\hbar^2/m_nAT}$, where $m_n$ corresponds to the
mass of the nucleon.

The mean value of any physical observable $\langle O\rangle $ is obtained by
weighing its value in each partition by the corresponding number of states
$w_f$:

\begin{equation}
\langle O \rangle_{\rm Micro} = \frac{\sum_f w_f O_f}{\sum_f w_f}\;.
\label{eq:avemc}
\end{equation}

\noindent
In particular, the average pressure $\langle P\rangle$ can be calculated
through:

\begin{equation}
P_f =-\frac{\partial F_f}{\partial V}=\frac{(M_f-1)T_f}{V_{\rm free}}
    +\frac{a_C^{f}}{3}\frac{V_0^{1/3}}{V^{4/3}}\;.
\label{eq:pmc}
\end{equation}

\noindent
The kinetic contribution is proportional to $M_f-1$, rather than to 
the total multiplicity, $M_f =\sum_{\{A,Z\}}N_{A,Z}$, as the
center of mass is kept at rest.
Even if it is not explicitly stated in the formulae, this condition is
consistently imposed in all the versions of the model.
Therefore, Eq.\ (\ref{eq:econs}), for instance, is actually modified to take
it into account.

Since, for large nuclei, the number of different fragmentation modes is
huge \cite{smm3}, we make a Monte Carlo sample of the most important
partitions.
As the sampling method does not select partitions with equal probability,
the statistical weight $w_f$ is corrected for this bias \cite{smm3}.
This procedure is also adopted in the canonical and in the isobaric
ensembles presented below.

Finally, a multiplicity dependent free volume has been introduced in
refs.\ \cite{smm2,smm3}:

\begin{equation}
\chi_f=\left[1+\frac{D}{2r_0A_0^{1/3}}(M^{1/3}_f-1)\right]^3-1\;,
\label{eq:muldep}
\end{equation}

\noindent
where $D=2.3r_0$.
This causes the breakup volume to fluctuate from
one partition to the other.
\end{subsection}

\begin{subsection}{The Canonical Ensemble}
\label{sec:canon}
If the breakup temperature of the system $T$ is fixed, instead of the
excitation energy of the source, besides its breakup volume, mass and atomic
numbers, the canonical ensemble is the best suited statistical treatment.
The partition function associated with a given fragmentation mode
reads:

\begin{equation}
Z_f^C(T,V)=\exp\left[\frac{-F_f(T,V)}{T}\right]\;.
\label{eq:zc}
\end{equation}

\noindent
Then, the average value of a physical observable is calculated through:

\begin{equation}
\langle O\rangle_{\rm C}=\frac{\sum_f Z_f^C O_f}{\sum_fZ_f^C}\;.
\label{eq:avec}
\end{equation}

Since the same essential ingredients are used in all the statistical ensembles,
the Helmholtz free energy $F$ is also given by Eq.\ (\ref{eq:helmfe}).
Therefore, Eq.\ (\ref{eq:pmc}) still holds for the pressure
in a given fragmentation mode.
The total energy of the system can be calculated though Eqs. (\ref{eq:fe})
and (\ref{eq:smc}), so that it is given by:

\begin{eqnarray}
E_f &=& \frac{3}{2}T(M_f-1)
+\sum_{\{AZ\}}N_{AZ}\left[-B_{AZ}+E^*_{AZ}(T)\right]\nonumber\\
&+&C_C^{f}\left(\frac{V_0}{V}\right)^{1/3}\;,
\label{eq:ecan}
\end{eqnarray}

\noindent
which now fluctuates from one fragmentation mode to the other since
the temperature is fixed.
The excitation energy is then given by:

\begin{equation}
E^*_f=E_f+B_{A_0Z_0}\;.
\label{eq:eexc}
\end{equation}

\end{subsection}

\begin{subsection}{The Isobaric Ensemble}
\label{sec:iso}
If one now fixes the pressure $P$, instead of the breakup volume,
the canonical ensemble can be modified to keep the pressure constant for
all microstates.
In this case, the partition function becomes:

\begin{equation}
Z_f^{\rm Iso}(T,P)=\int_{V_0}^\infty\,dV\;Z_f^C(T,V)
                  \exp\left(-\frac{PV}{T}\right)\;,
\label{eq:ziso}
\end{equation}

\noindent
which, by using Eqs. (\ref{eq:helmfe}-\ref{eq:fec}), can be rewritten as:

\begin{equation}
Z_f^{\rm Iso}(T,P)=\tilde Z_f(T,P)I_f(T,P)\;.
\label{eq:ziso2}
\end{equation}

\noindent
In the equation above, we have defined:

\begin{equation}
I_f(T,P)=\int_0^\infty d\chi\, \chi^{M_f-1}
\exp\left(-\frac{Q_f(\chi,P)}{T}\right)\;,
\label{eq:intiso}
\end{equation}

\begin{equation}
Q_f(\chi,P)=PV_0\chi
+\frac{a_C^f}{(1+\chi)^{1/3}}\;,
\label{eq:qi}
\end{equation}

\begin{equation}
\tilde Z_f(T,P)=\frac{V_0^{M_f}}{\lambda_T^{3(M_f-1)}}
                \exp\left(-\frac{PV_0}{T}\right)\prod_{\{A,Z\}}\Gamma_{A,Z}\;,
\label{eq:ztiso}
\end{equation}

\noindent
and

\begin{equation}
\Gamma_{A,Z}=\frac{(g_{A,Z}A^{3/2})^{N_{A,Z}}}{N_{A,Z}!}
                \exp\left[\frac{N_{A,Z}}{T}[B_{A,Z}-F^*_{A,Z}(T)]\right]\;.
\label{eq:ztiso2}
\end{equation}

\noindent

Similarly to the former case, average values are obtained through:

\begin{equation}
\langle O\rangle_{\rm iso}=\frac{\sum_f Z_f^{\rm Iso} O_f}
                                {\sum_f Z_f^{\rm Iso}}\;.
\label{eq:aveiso}
\end{equation}

The appropriate thermodynamical function for the isobaric ensemble is
the Gibbs free energy $G$:

\begin{equation}
G=F+PV\;,
\label{eq:gibbs}
\end{equation}

\noindent
which is related to the partition function by:

\begin{equation}
G_f(T,P)=-T\log\left[Z_f^{\rm Iso}(T,P)\right]\;.
\label{eq:gibbsziso}
\end{equation}

\noindent
From it, one may obtain the average volume of the system, for a given
fragmentation mode:

\begin{equation}
\overline{V}^f=\frac{\partial G_f}{\partial P}
              =V_0\frac{I_f^V(T,P)}{I_f(T,P)}\;,
\label{eq:avevol}
\end{equation}

\noindent
where

\begin{equation}
I_f^V(T,P)=\int_0^\infty d\chi\, (1+\chi)\chi^{M_f-1}
\exp\left(-\frac{Q_f(\chi,P)}{T}\right)\;.
\label{eq:ivol}
\end{equation}

\noindent
The energy of the system corresponding to the breakup channel $f$ can then be
written as:

\begin{eqnarray}
E_f&=&\frac{\partial {\;}}{\partial (1/T)}\left(\frac{G_f}{T}\right)
   -P \overline{V}^f\nonumber\\
&=&PV_0\left[1+\frac{I_f^E(P,T)}{I_f(P,T)}-\frac{I_f^V(P,T)}{I_f(P,T)}\right]+\frac{3}{2}(M_f-1)
\nonumber\\
&+&\sum_{\{A,Z\}}N_{A,Z}\left[-B_{A,Z}+E^*_{A,Z}(T)\right]\;,
\label{eq:eiso}
\end{eqnarray}

\noindent
so that

\begin{equation}
E^*_f=E_f+B_{A_0Z_0}\;,
\label{eq:eexiso}
\end{equation}

\noindent
where

\begin{eqnarray}
I_f^E(T,P)=\int_0^\infty
&&\left[\chi+\frac{a_C^f}{PV_0}\frac{1}{(1+\chi)^{1/3}}\right]\chi^{M_f-1}
\nonumber\\
&&\exp\left(-\frac{Q_f(\chi,P)}{T}\right)\,d\chi\;.
\label{eq:inte}
\end{eqnarray}

\noindent
Owing to the Coulomb factor $(1+\chi)^{-1/3}$, all the integrals above
have to be evaluated numerically for each fragmentation mode.
Since the integrands are bell shaped, the gaussian quadrature method is very
efficient.

Finally, from the above expressions, the entropy can be evalutated
from the relation:

\begin{equation}
S_f=\frac{E_f+P\overline{V}^f-G_f}{T}\;.
\label{eq:entropy}
\end{equation}

\end{subsection}

\end{section}

\begin{section}{Comparison among the ensembles}
\label{sec:comparison}
All the example calculations in this work are carried out for a $A=168$ and
$Z=75$ system.
The same breakup temperature is used in the isobaric and in the
canonical calculations.
We fixed the pressure, in the isobaric ensemble, at $p=0.114$~MeV/fm$^3$.
Although our conclusions are not qualitatively affected by this particular
choice, the results below show that this is a reasonable value and lies
within the expected range for this system \cite{msulong2003}.
Since the micro-canonical ensemble requires the energy as input,
we use the average excitation energy obtained in the isobaric
ensemble, calculated through Eqs.\ (\ref{eq:eiso}-\ref{eq:eexiso})
and (\ref{eq:aveiso}), for a given temperature.
The breakup volume, which is a free parameter for the canonical and
micro-canonical ensembles, is chosen in different ways, as explained below.

\begin{figure}[ht]
\includegraphics[angle=0,totalheight=7.0cm]{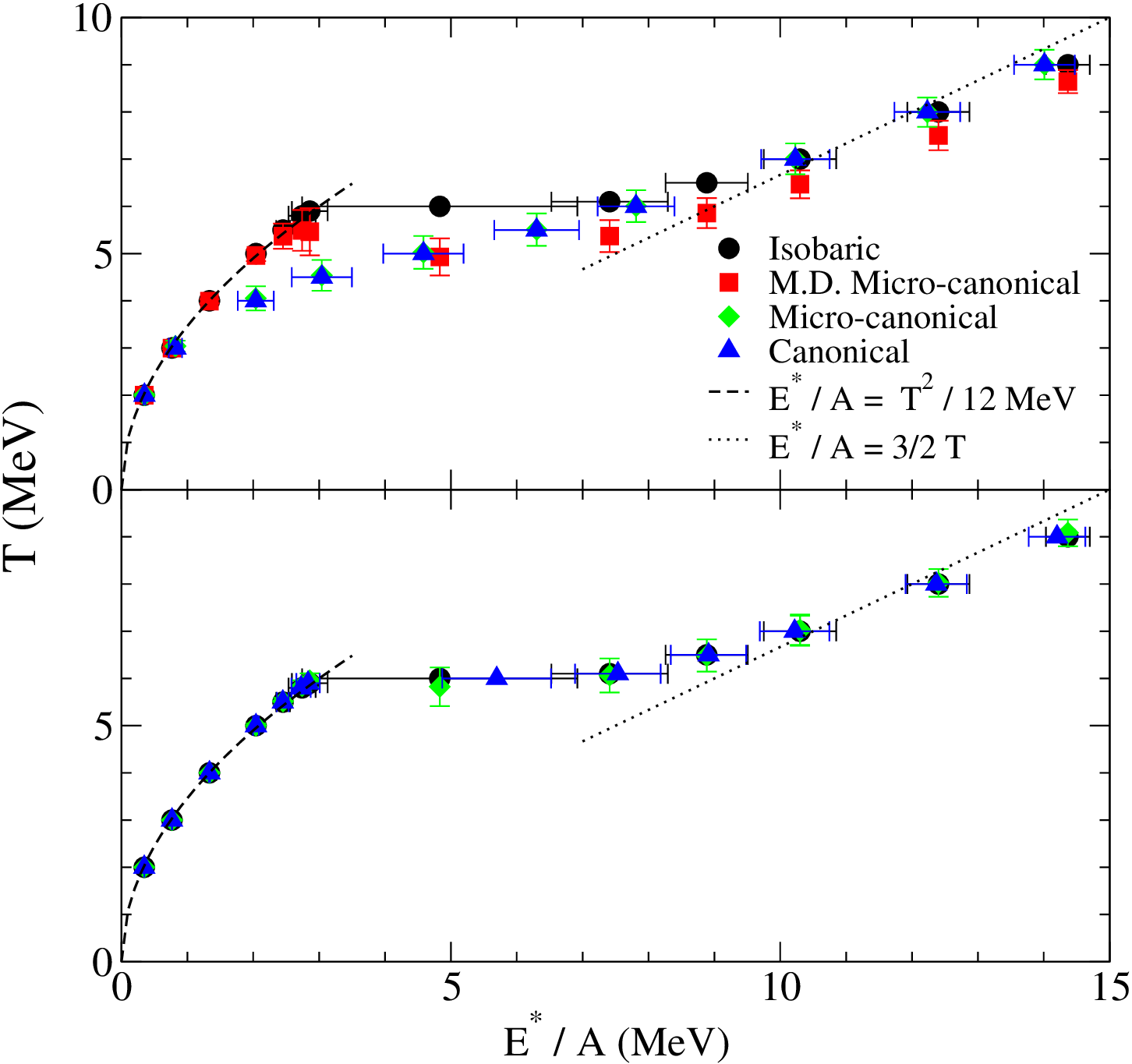}
\caption{Caloric curve predicted by the different ensembles.
Upper panel: the breakup volume is kept constant for all energies for
the canonical and the standard micro-canonical ensembles.
Lower panel: the average breakup volume predicted by the isobaric calculation
is used as input for the other ensembles.
For details, see text.}
\label{fig:cc}
\end{figure}

Before we present the model results for the different ensembles,
it is important to stress that the error bars in all the figures
correspond to the standard deviations of the distributions, rather than
to statistical fluctuations.
Indeed, all the curves show a smooth behavior, even when the error bars
are appreciably large.
We included this information in all the pictures so as to provide more
information about the distributions instead of only mean values.

We start by analysing the caloric curve predicted by the isobaric ensemble,
which is depicted by the circles in Fig.\ \ref{fig:cc}.
It shows that, at low temperatures ($T < 6$~MeV), the caloric curve
follows the standard Fermi gas expression (represented by the
dashed line in this picture), in agreement with previous
calculations \cite{smm2}.
However, a sudden change takes place at $T=6$~MeV and one observes a
fairly wide plateau, which signals a liquid-gas phase transition.
For higher temperatures, the system tends to behave as a free gas, as
also observed in ref. \cite{smm2}.
We have checked that the critical temperature depends on the pressure, but
the essential features of the process remain the same.

It is well known that SMM predicts the existence of a plateau in the
caloric curve only if the breakup volume, for a fixed excitation energy,
is allowed to vary from one fragmentation mode to the other
\cite{smm2,bondorfcc,msulong2003,isocc}.
Indeed, we also show, in the upper panel of Fig.\ \ref{fig:cc},
the results obtained using two versions of SMM.
The one for which the breakup volume is determined, for each fragmentation
mode, by Eqs.\ (\ref{eq:vfree}) and (\ref{eq:muldep}), is labelled
``M.D.~micro-canonical'' (Multiplicity Dependent), to distinguish it from
the ``standard'' micro-canonical version (with no additional labels)
in which the breakup volume is fixed for all partitions at a given
excitation energy.
The results clearly show that the temperature increases monotonously if the
breakup volume is kept fixed for all energies (diamonds) although a behavior
similar to that predicted by the isobaric ensemble is observed in the other
case (squares).
In all calculations in which the breakup volume is the same for all energies,
we adopt $V/V_0=3$, for both micro-canonical and canonical ensembles.

The predictions made by the canonical version of SMM are represented by the
triangles in the upper panel of Fig.\ \ref{fig:cc}.
An excelent agreement between the canonical and the microcanonical
ensembles is obtained, showing the consistency of the model, as long as the
relevent macroscopic quantites, such as breakup volume, average
temperature/excitation energy, and pressure, are similar.

This expectation is reinforced through the comparison among the caloric curves
predicted by the isobaric, standard micro-canonical, and canonical ensembles
shown in the lower panel of Fig.\ \ref{fig:cc}.
In this case, the breakup volume for the latter two ensembles, for
each temperature/excitation energy, is taken to be
the average value obtained in the isobaric calculation.
It is clear that all the curves collapse into a single one.
This implies that the most appropriate statistical scenario for nuclear
multifragmentation may not be singled out by examining average values only.

\begin{figure}[ht]
\includegraphics[angle=0,totalheight=7.0cm]{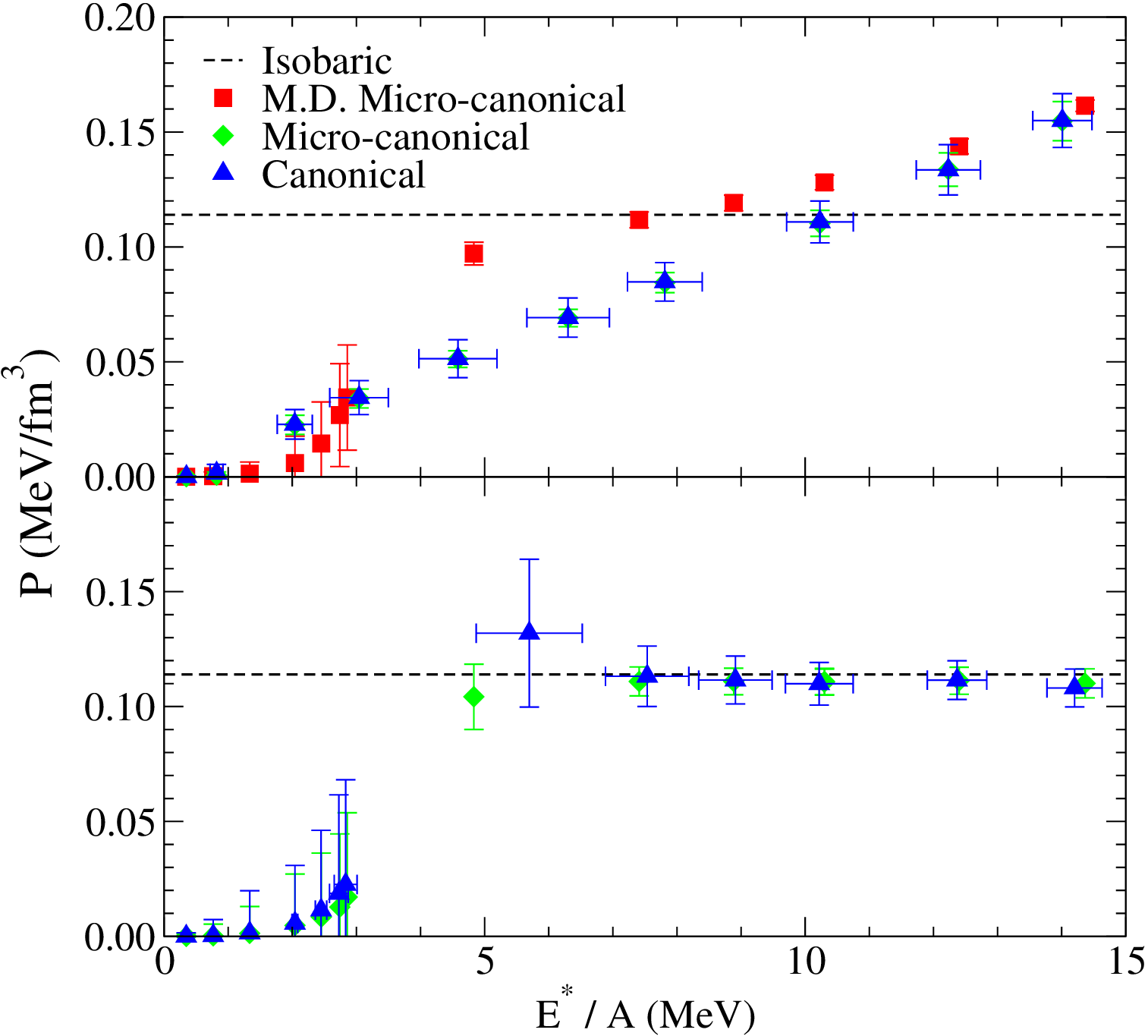}
\caption{Average pressure predicted by the different ensembles as a
function of the excitation energy.
In the upper panel the breakup volume is kept constant for all
energies, except for the M.D.~micro-canonical version.
The results displayed in the lower panel are obtained by using the
average breakup volume predicted by the isobaric ensemble
as input to the canonical and micro-canonical calculations.}
\label{fig:press}
\end{figure}

To further clarify the aspects discussed above, we now investigate the
average pressure predicted by the different versions of the model.
The results shown in Fig.\ \ref{fig:press} are calculated from
Eqs.\ (\ref{eq:pmc})-(\ref{eq:avemc}) and Eqs.\ (\ref{eq:pmc})-(\ref{eq:avec}),
for the micro-canonical and canonical ensembles, respectively.
A steady increase of the pressure as a function of the excitation energy
is observed, in the upper panel of this figure, for the standard micro-canonical
and the canonical calculations.
Since the breakup volume is the same for all excitation energies,
this behavior is quite reasonable on physical grounds.
On the other hand, fluctuations of the breakup volume, from one fragmentation
mode to the other, introduced by Eqs.\ (\ref{eq:vfree}) and (\ref{eq:muldep}),
lead to a weak energy dependence in the plateau region, as revealed by the
results obtained employing the corresponding micro-canonical model, represented
by the squares in the upper panel of this picture.

The role played by the volume change is emphasized by the results shown
in the lower panel of the same picture.
In this case, the breakup volume for the canonical and micro-canonical
calculations is chosen so that it corresponds to the average value
obtained in the isobaric ensemble for each temperature.
The pressure now stays fairly constant for $E^*/A> 5$~MeV.
For lower excitation energies, it decreases owing to the influence of
the large remnant, which, in this energy range, is almost always present.

The prediction of a plateau in the caloric curve and the similarity
of the average pressure, in this region, found in the isobaric and
the M.D.~micro-canonical calculations, seem to indicate that the {\it ad hoc}
multiplicity dependence, in the latter, Eqs.\ (\ref{eq:vfree})
and (\ref{eq:muldep}), leads to similar breakup volumes in both statistical
ensembles.
This is illustrated in Fig.\ \ref{fig:volume}, which displays the
average breakup volume obtained with these two versions of the model.
As expected, the predictions are very similar, although the micro-canonical
ensemble gives smaller average breakup volumes at high energies.
This explains why the average pressure increases for $E^* / A > 10$~MeV.
A striking feature of the isobaric ensemble is the linear energy dependence
of the breakup volume after the onset of multifragment emission.
This is expected to happen for a gas with no internal degrees of freedom but,
particularly at the plateau, the contribution from complex fragments is not
negligible.

\

\begin{figure}[ht]
\includegraphics[angle=0,totalheight=5.5cm]{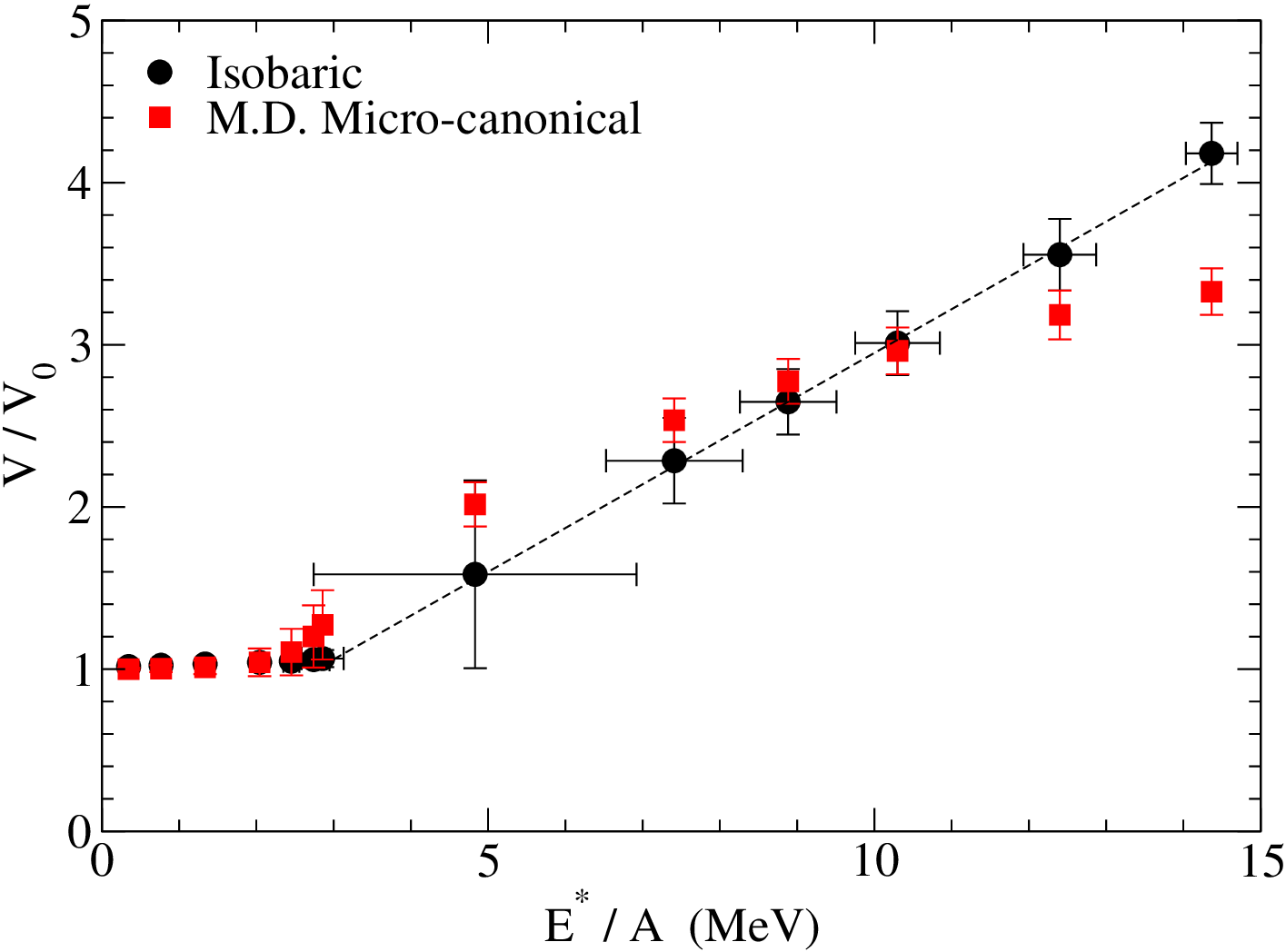}
\caption{Average breakup volume as a function of the excitation energy
predicted by the isobaric and the micro-canonical ensembles.
The dashed line just emphasizes the fairly linear dependence of
the breakup volume on the excitation energy in the isobaric ensemble.
For details see text.}
\label{fig:volume}
\end{figure}

Although the compatibility among the different versions of the model
can be understood, some qualitative aspects still depend strongly upon
the version one uses.
This is demonstrated by the plot of the entropy as a function of the
temperature in Fig.\ \ref{fig:entropy}.
In the upper panel, the canonical and the standard microcanonical
calculations were carried out at fixed breakup volume for all temperatures.
One sees that the entropy rises smoothly as the excitation energy increases.
However, a backbending is observed in the results obtained with the
M.D.~microcanonical version.
This led the authors of ref.\ \cite{smm2} to predict negative heat capacities
in the plateau region.
However, our results (Fig.\ \ref{fig:press}) show that the pressure is only
approximately constant in the microcanonical calculations.
Thus, one should be careful when drawing conclusions about $C_p$ from the
entropy plot within this microcanonical
calculation, since the microstates are not subject to a constant pressure.

\begin{figure}[ht]
\includegraphics[angle=0,totalheight=7.0cm]{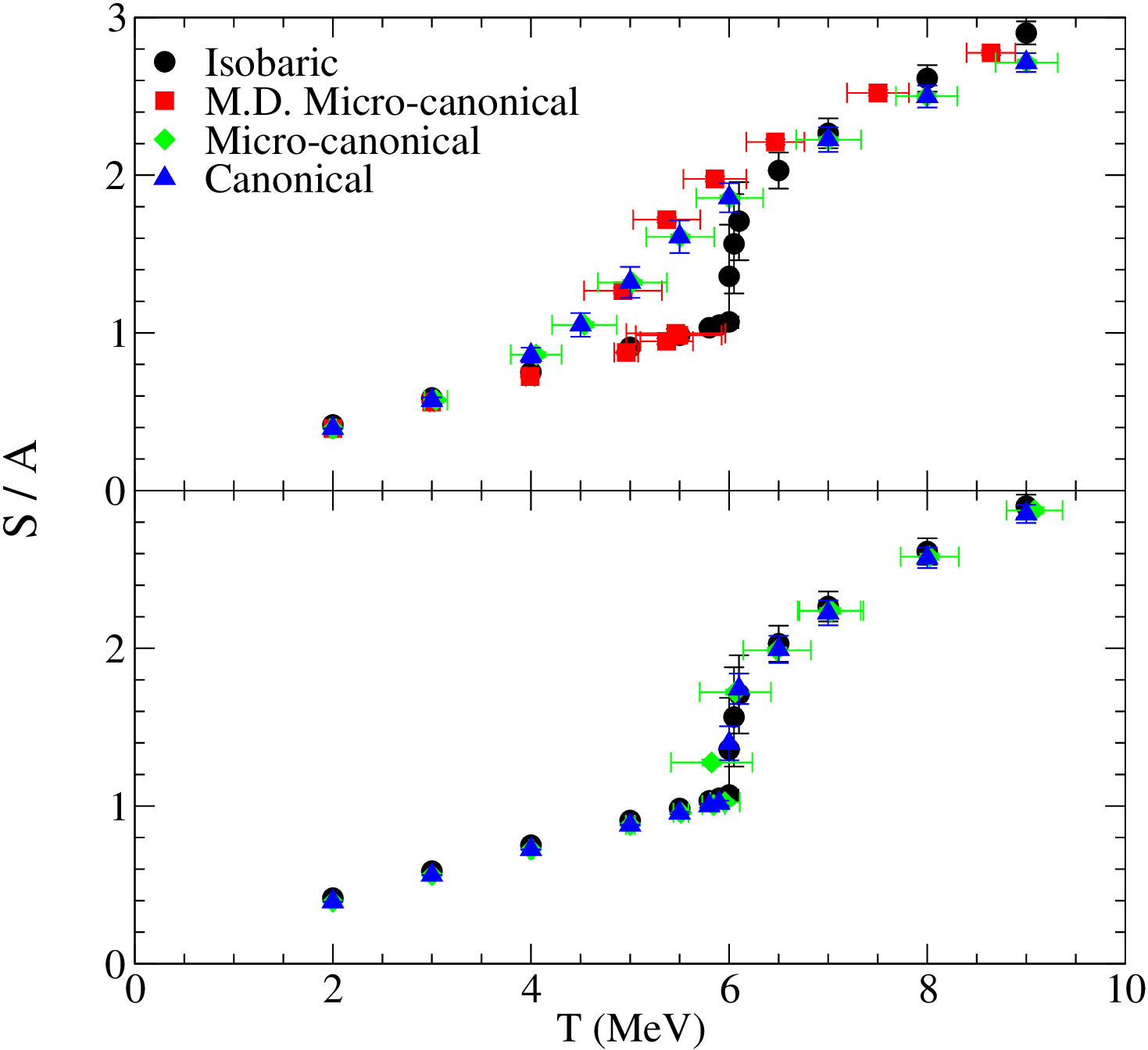}
\caption{Entropy per nucleon as a function of the temperature.
Results corresponding to the canonical and the standard microcanonical
ensembles are obtained at a constant volume for all temperatures in the upper
panel, whereas the average values given by the isobaric ensemble were
used as input in the lower panel. For details see text.}
\label{fig:entropy}
\end{figure}

On the other hand, from the results obtained with the isobaric ensemble,
depicted by the circles in this picture, one
can certainly assert that $C_p$ is very large, but always positive, in the
plateau region.
Nevertheless, the consistency of the model is not affected by this
disagreement as is shown in the lower panel of Fig.\ \ref{fig:entropy},
in which the breakup volume for the canonical and the microcanonical
calculations are taken from the average values predicted by the isobaric
ensemble at each temperature.
Also in this case, a very good agreement among the ensembles is found.
The microcanonical calculations still seem to predict negative $C_p$ in
a narrow temperature region, although one should keep in mind the
remarks above.
The relevant point here is that all ensembles make very similar predictions
under equivalent conditions.

\begin{figure}[ht]
\includegraphics[angle=0,totalheight=6.0cm]{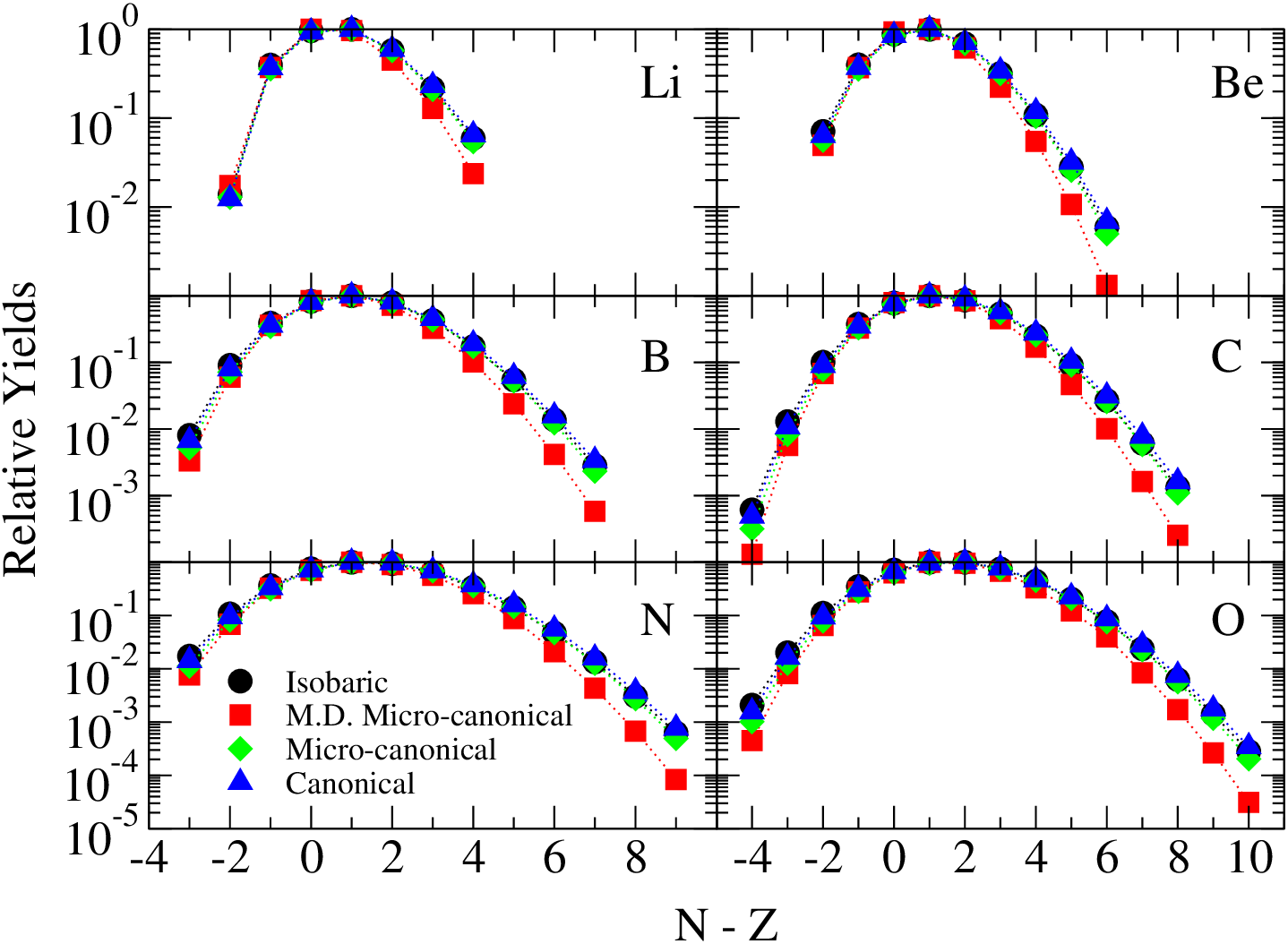}
\caption{Isotopic distributions predicted by the different ensembles.
The yields are normalized relative to the maximum for each $Z$ value.}
\label{fig:isotopes}
\end{figure}

Even though it is difficult to determine experimentally the heat capacity,
other observables are very sensitive to the statistical assumptions
and could, therefore, help to decide which picture is more
convenient to describe the multiframent emission mechanism.

For instance, in Fig.\ \ref{fig:isotopes}, we show the isotopic distributions
for a few selected cases obtained using the different versions of the model.
One notices that the isobaric, canonical, and the standard microcanonical
ensembles give very similar isotopic distributions.
As above, the breakup volume for the canonical and the microcanonical
models is taken from the isobaric calculations.
In contrast, the isotopic distributions predicted by the M.D.~microcanonical
are systematically narrower.
Since precise data are avaliable for many reactions \cite{msuIsotopes},
important conclusions may be drawn from the study of this observable.

\

\

\begin{figure}[ht]
\includegraphics[angle=0,totalheight=5.5cm]{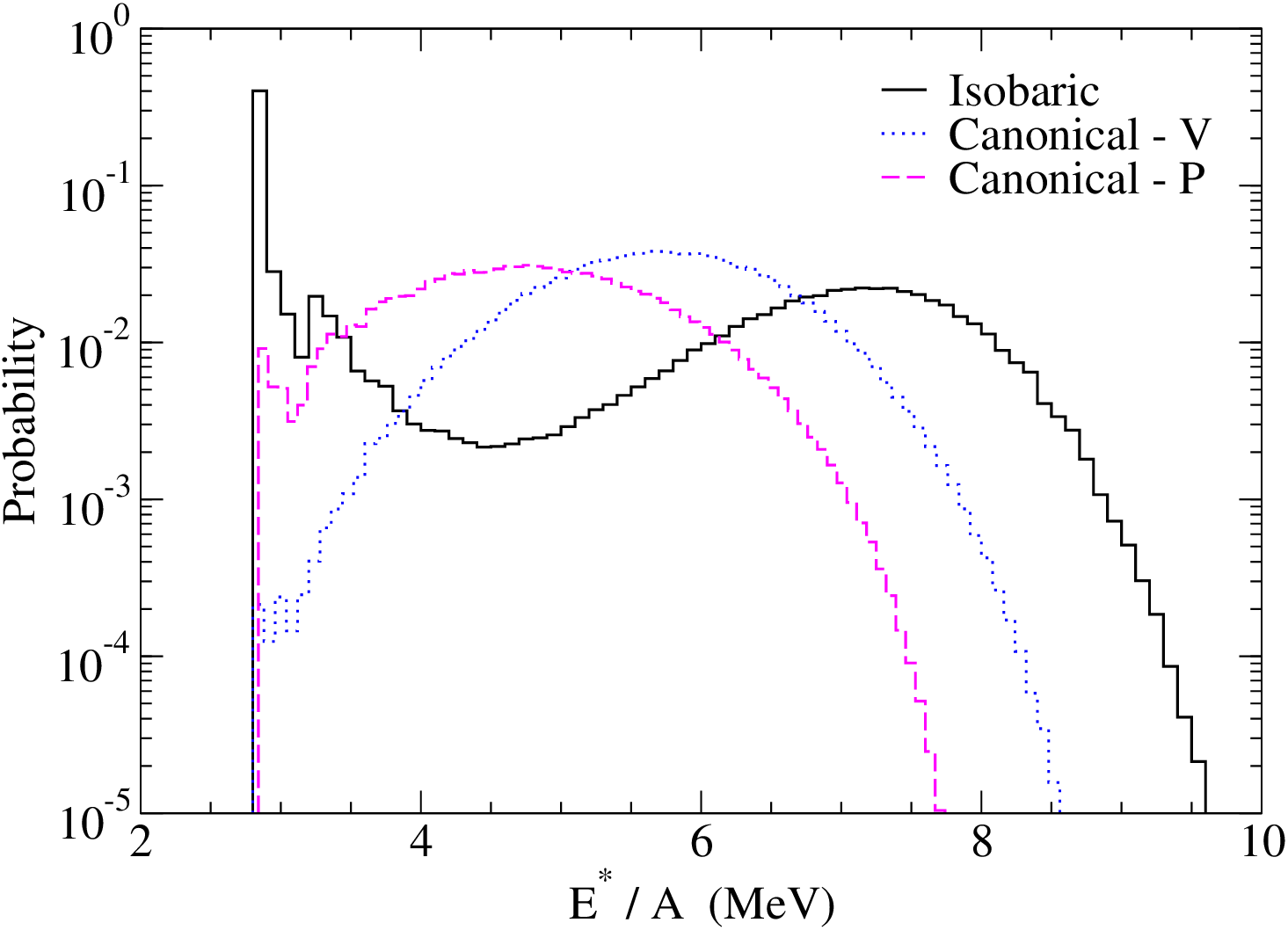}
\caption{Energy distribution obtained with the isobaric and canonical
ensembles calculated at $T=6$~MeV. For details see text.}
\label{fig:he}
\end{figure}

Another key observable is the excitation energy distribution, which
is shown in Fig.\ \ref{fig:he} for the isobaric and canonical ensembles
at $T=6$~MeV.
As can be noticed, it turns out to be quite sensitive to the statistical
assumptions.
The canonical calculations have been performed in two different ways.
In one case, the breakup volume corresponds to the average value obtained
in the isobaric calculation and is labelled canonical-V.
In the second canonical calculation, the breakup volume is chosen so that
it gives the average pressure $p=0.114$~MeV/fm$^3$, and labelled
canonical-P.
These two canonical calculations agree qualitatively, as both give bell
shaped distributions, which essentially differ by the mean values.
On the other hand, when all the microstates are at the same pressure,
the energy distribution has a narrow peak at low excitation energy and a
smeared bump at high excitation energies, with a large gap between them.
The low energy part of the distribution correspond to the presence of a
large remnant (the liquid phase) in the breakup channel, whereas essentially
light fragments are found at high energies (the gas phase).

This is illustrated in Fig.\ \ref{fig:ha}, which shows the results obtained
with the isobaric and canonical ensembles.
The canonical calculations are done in the manner just described.
One should notice that the tendency to produce a large remnant in the
breakup channel seems to be a pressure efffect as the canonical calculation
which mimics the isobaric ensemble appears to produce heavy fragments
and even shows a small bump near $A_0$.
This behavior is not seen in the other canonical calculation.

\

\

\begin{figure}[ht]
\includegraphics[angle=0,totalheight=5.5cm]{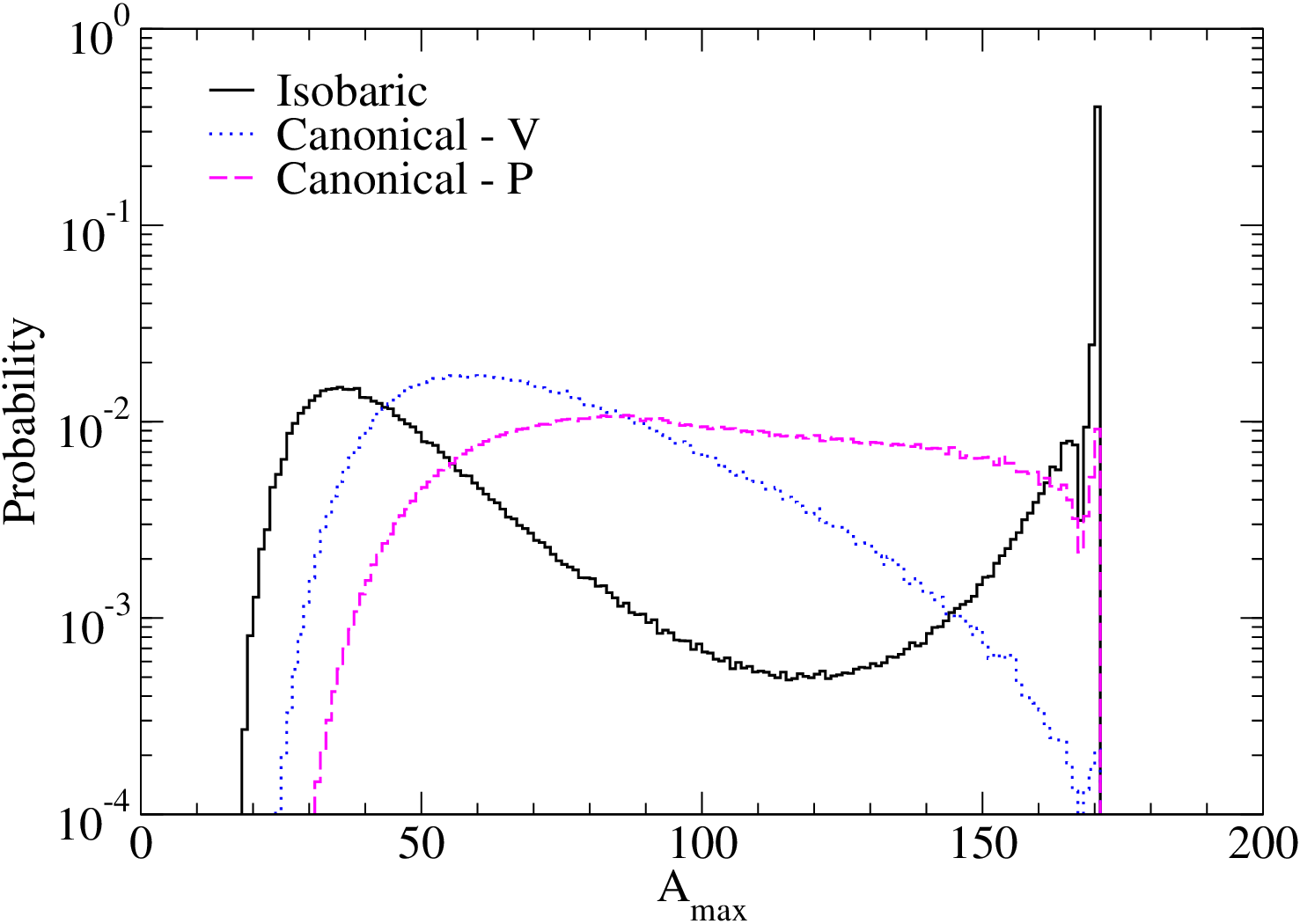}
\caption{Distribution of the largest fragment within each fragmentation
mode obtained with the isobaric and canonical ensembles at $T=6$MeV.}
\label{fig:ha}
\end{figure}

\end{section}

\begin{section}{Concluding remarks}
\label{sec:conclusion}
We have examined the sensitivity of the predictions of the SMM to its
statistical assumptions.
We found that the microcanonical, canonical, and isobaric implementations
of SMM predict very similar caloric curves, and other (average) physical
observables, provided macroscopic variables, such as temperature, excitation
energy, and breakup volume, are the same in all versions of the model.

Although this is not the main point of this work, it is worth reporting that
we found large values of $C_p$ at the plateau region of the caloric curve,
but no indication of negative heat capacities in our isobaric calculations, in
agreement with ref.\ \cite{MorettoNegativeHeatCapacity2002}.
We have concentrated on the $A_0=168$, $Z_0=75$ system, but we have also checked
that this conclusion still holds for lighter systems, such as the
$^{108}$Ag nucleus, in contrast with other calculations
\cite{smm2,dasGuptaCp,theNatureOfPhaseTransition2001,Elliott2000,
IsobaricShlomo2004}.
More specifically, we observed negative $C_p$ only in the
microcanonical calculations, as already reported in refs.
\cite{Chomaz2000,GrossPhysRep1997,smm2}.
Nevertheless, the predictions made by the
isobaric ensemble should be more reliable since it is the only statistical
ensemble in which all the microcospic states are subject to the same
pressure, whereas, in the other cases, this condition if fulfilled only
on the average.
We should also mention that we did not investigate the sensitivity of this
conclusion to the model parameters related to the binding and
excitation energy of the source.
As reported in ref.\ \cite{dasGuptaCp} these parameters may influence the
mass region in which negative $C_p$ values are obtained in the statistical
calculations.

We would like to stress the point that the distributions of physical
observables, rather than their mean values, seem to be fairly sensitive
to the statistical assumptions.
The relevance of this conclusion goes beyond the fact that the equivalence of
the statistical ensembles should hold only in the thermodynamical limit
\cite{ChomazLG2005}.
Indeed, some predictions made upon different assumptions agree very well,
as long as the most relevant input global parameters do not differ appreciably.
Strictly speeking, one may distinguish among the different
calculations.
However, we belive that experiments cannot be precise enough to single out one
of them only by examining average values.
Therefore, although nuclear systems are small and nuclear reactions cannot be
fine tuned, the distributions of some physical observables should bear 
clear signatures related to the underlying statistical assumptions.
Then, the analysis of such distributions may be valuable in establishing
an appropriate scenario for the freeze-out stage in nuclear multifragmentation.

\end{section}

\

\begin{acknowledgments}
We would like to acknowledge CNPq, FAPERJ, and the MCT/FINEP/CNPq (PRONEX)
program, under contract \#41.96.0886.00, for partial financial support.
\end{acknowledgments}

\bibliography{isobaric}

\end{document}